\documentclass[fleqn,usenatbib]{mnras}

\usepackage[T1]{fontenc}

\DeclareRobustCommand{\VAN}[3]{#2}
\let\VANthebibliography\thebibliography
\def\thebibliography{\DeclareRobustCommand{\VAN}[3]{##3}\VANthebibliography}

\usepackage{graphicx}	
\usepackage{amsmath}	
\usepackage{amssymb}	

\usepackage{newtxtext,newtxmath}

\title[Flux and colour variability in blazars using ZTF]{Optical Flux and Colour Variability of Blazars in the ZTF Survey}

\author[Negi et al.]{
Vibhore Negi$^{1,2}$,\thanks{E-mail:vibhore.negi18@gmail.com}
Ravi Joshi$^{3,4}$,
Krishan Chand$^{1,8}$,
Hum Chand$^{5,1}$,
Paul Wiita$^{6}$,
Luis C. Ho$^{4,7}$,
\newauthor {  and Ravi S. Singh$^{2}$}
\\\\
$^{1}$Aryabhatta Research Institute of Observational Sciences (ARIES), Manora Peak, Nainital, 263002, India\\
$^{2}$Department of Physics, Deen Dayal Upadhyaya Gorakhpur University, Gorakhpur, 273009, India\\
$^{3}$Indian Institute of Astrophysics, Koramangla, Bangalore, 560034, India\\
$^4$Kavli Institute for Astronomy and Astrophysics, Peking University, Beijing, 100871, China\\
$^{5}$Department of Physics and Astronomical Sciences, Central University of Himachal Pradesh (CUHP), Dharamshala, 176215, India\\
$^{6}$Department of Physics, The College of New Jersey, PO Box 7718, Ewing, NJ, 08628-0718, USA\\
$^{7}$Department of Astronomy, School of Physics, Peking University, Beĳing, 100871, China\\
$^{8}$Department of Physics, Soban Singh Jeena University, Almora, 263601, India
}

\date{Accepted XXX. Received YYY; in original form ZZZ}

\pubyear{2021}

\begin{document}
\label{firstpage}
\pagerange{\pageref{firstpage}--\pageref{lastpage}}
\maketitle

\begin{abstract}
We investigate the temporal and colour variability of 897 blazars, comprising 455 BL Lacertae objects (BL Lacs) and 442 Flat Spectrum Radio Quasars (FSRQs), selected from the Roma-BZCAT catalogue, using the multi-band light curves from the Zwicky Transient Facility (ZTF DR6) survey. Assessing the colour variability characteristics over $\sim$2\ year timescales, we found that 18.5 per cent (84 out of 455) BL Lacs showed a stronger bluer when brighter (BWB) trend, whereas 9.0 per cent (41 out of 455) showed a redder when brighter (RWB) trend. The majority (70 per cent) of the BL Lacs showing RWB are host galaxy dominated. For the FSRQ subclass, 10.2 per cent (45 out of 442) objects showed a strong BWB trend and 17.6 per cent (78 out of 442) showed a strong RWB trend. Hence we find that  BL Lacs more commonly follow a BWB trend than do FSRQs. This can be attributed to the more dominant jet emission in the case of BL Lacs and the contribution of thermal emission from the accretion disc for FSRQs. In analysing the colour behaviour on shorter time windows, we find many blazars evince shorter partial trends of BWB or RWB nature (or occasionally both). Some of such complex colour behaviours observed in the colour-magnitude diagrams of the blazars may result from transitions between the jet-dominated state to the disc-dominated state and vice versa.

\end{abstract}

\begin{keywords}
galaxies: active --- BL Lacertae objects: general --- quasars: general ---   galaxies: jets
\end{keywords}



\section{Introduction}

Blazars are a special subclass of active galactic nuclei (AGN), possessing a  relativistic jet directed close to the observer's line of sight  \citep{1978bllo.conf..328B,1995PASP..107..803U}. The blazars emit in a very wide range of wavelengths from radio to gamma-ray bands and show violent variability on diverse timescales ranging from minutes to years \citep[e.g.,][]{1992ApJ...386..473H}. Their spectral energy distributions (SEDs) are characterised by non-thermal continuum spectra consisting of two components: a low energy component peaking anywhere from infrared to X-ray frequencies and a high energy component peaking at frequencies from hard X-rays to TeV gamma-rays. The low energy component is attributed to the synchrotron radiation emitted by relativistic electrons in the jets \citep{1982ApJ...253...38U}. While the origin of the high energy component is not fully understood, it is commonly thought to be due to the inverse-Compton scattering of soft photons by the same relativistic electrons \citep[e.g.,][]{1992A&A...256L..27D, 1994ApJ...421..153S}.
\par
Blazars are broadly classified into two categories: BL Lacertae objects (BL Lacs) and flat spectrum radio quasars (FSRQs) \citep{1980ARA&A..18..321A}, based on the rest-frame equivalent widths (EWs) in their spectra being $<$ 5 \AA\ and $>$ 5\AA, respectively \citep[e.g.,][]{2011MNRAS.414.2674G}. BL Lacs are further classified into three subcategories depending on the peak frequency of their synchrotron emission: high-, intermediate-, and low-energy peaked BL Lac objects (HBL, IBL, and LBL). The peak frequencies for HBLs, IBLs, and LBLs lie higher than, around, and lower than the optical band, respectively. 
\par
The study of blazar variability is an important tool that provides valuable information about their nature. The optical variability of blazars has  been a topic of intensive research since their discovery as it affords a unique perspective on accretion disc and jet physics \citep{1995ARA&A..33..163W,2000ApJS..127...11G,2004A&A...421..103V,2008AJ....135.1384G,2010MNRAS.404.1992R,2011PASJ...63..639I,2011A&A...534A..59G,2012ApJ...756...13B,2016Ap&SS.361..345M}. In particular, optical brightness variations in blazars are often accompanied by spectral variations that can be revealed by the colour-magnitude (or spectral index-magnitude) correlations \citep[e.g.,][]{2003A&A...400..487C}. The analysis of spectral variability is of special interest in the sense that it might constrain the origin of corresponding brightness variations, even when restricted to the optical band \citep{2003ApJ...590..123V}.
\par
Past studies of blazars in the optical band have detected several colour-magnitude correlation patterns. It has been claimed that for BL Lacs, a bluer when brighter (BWB) trend, i.e., colour becoming bluer with increasing brightness, is more commonly seen. This trend possibly arises because electrons accelerated at the front of the shock lose energy while propagating away from it, and since the higher energy electrons lose energy faster through synchrotron cooling, thus making the high frequency bands more variable \citep[e.g.,][]{1998A&A...333..452K,2019MNRAS.488.4093A}. However, an opposite redder when brighter (RWB) trend may be more common in the case of FSRQs \citep[e.g.,][and references therein]{2006A&A...450...39G,2010MNRAS.404.1992R}, likely due to an increase in the relative contribution of the more variable beamed jet emission to the less variable, but bluer, accretion disc emission. At the same time, some studies of a few sources have reported complicated colour-magnitude relations: a stable-when-brighter (SWB) trend, a BWB trend during a period of time but an RWB trend during another period of time, or a BWB/RWB trend not strong enough to make any such claim \citep{2011PASJ...63..639I,2012ApJ...756...13B,2015RAA....15.1784Z,2017ApJ...844..107I}.
However, these studies have been performed by intensively monitoring individual objects or a few tens of objects and are luminosity biased, leading to heterogeneous observations and small sample sizes. Consequently, despite these efforts, and because of the expansive observing times required for them, a clear picture of the colour behaviour of the blazars has not been established. However, with the advent of several new time-domain surveys offering good cadences, it has become possible to investigate the colour variability of the blazar populations in an unbiased manner. \\

In this work, we investigate for the first time whether there is a universality in the colour behaviour of the blazars with the largest homogeneous sample taken to date. We have made use of Zwicky transient facility \citep[ZTF,][]{2019PASP..131a8002B} which scans the northern sky in $g$, $r$, and $i$ bands with an average three days cadence using a 47 deg$^{2}$ wide-field imager mounted on a 48-inch Schmidt telescope on Mount Palomar. The paper is organised as follows. In section 2, we describe the catalogues used to get our source list, the data used, and the procedure to define the final sample for our study. Section 3 describes the data analysis and the results obtained from it. In section 4, we discuss these results and the key conclusions are presented in section 5.

\section{Data and Sample selection}
\label{sec:Sample} 
We employ the latest version (5th edition) of the Roma-BZCAT catalogue \citep{2015Ap&SS.357...75M}, the most complete list of all the blazars detected in multi-frequency surveys to date. It consists of a  total of 3561 blazars, including 1425 BL Lacs, with 92 of them being BL Lac candidates, 1909 FSRQs, and 227 blazars of uncertain type. Additionally, out of the confirmed BL Lacs, 274  have a spectral energy distribution (SED) with a significant dominance of the galaxy's emission over that of the nucleus. In order to have a clear picture of the characteristics of BL Lacs and FSRQs, we have excluded the 227 blazars listed as uncertain type and the 92 BL Lac candidates in BZCAT that led us to a total of 3242 sources. 

Next, we searched for the point spread function (PSF) fit based light curves from the 6th ZTF public data release\footnote{\url{https://www.ztf.caltech.edu}} \citep[see also,][]{2019PASP..131a8003M} for all the objects centred at the target position within an angular separation of 1.5$\arcsec$ radius. Since the $i$ band observations were not well sampled for most of our targets, we first compiled a preliminary sample of 2471 sources observed in both $g$ and $r$ bands.
Note that ZTF treats the light curves observed in a particular field, filter, and CCD-quadrant independently and assigns a different observation ID to the source observed in different combinations of the three. The ZTF quadrants are calibrated independently, and thus combining light curves from different fields and CCD-quadrants (but same filter) can produce spurious variability \citep{2021AJ....161..267V}. Therefore, to avoid the spurious variability originating from these different calibrations, we only take the light curve corresponding to the observation ID with the maximum number of data points. 
The ZTF DR6 also provides a quality score of the observed data obtained in good observing conditions\footnote{\url{https://www.ztf.caltech.edu/page/dr6}}, as catflags score = 0. Therefore, we rejected all the photometric points with a non-zero catflags score. Given that photometric outliers and abnormally large errors can significantly affect the variability measurements, we removed the outliers by applying a 3$\sigma$ clipping on the whole data set and also excluded the poor quality data points with large uncertainties, taken to be those exceeding 10 per cent in magnitude. \\

\hskip-10.0cm
\begin{table}
	\centering
	\setlength\tabcolsep{3pt}
	\caption{Basic parameters of the blazar sample used in our study. }
	\label{tab:sample_table}
	\begin{tabular}{cccccccc} 
		\hline
		Object & RA (J2000) & DEC (J2000) & Type & $R_{mag}$ & z\\
		&  (hh:mm:ss) &   (dd:mm:ss) &   &  (ROMABZ)  &   \\
		\hline
		J161706+410647 & 16:17:06.31 & +41:06:47.01 & BL Lac & 17.7 & 0.267\\
		J163515+380804 & 16:35:15.48 & +38:08:04.48 & FSRQ & 17.6 & 1.814\\
		J171613+683638 & 17:16:13.93 & +68:36:38.69 & FSRQ & 17.8 & 0.777\\
		J172727+453039 & 17:27:27.65 & +45:30:39.70 & FSRQ & 18.5 & 0.717\\
		J175728+552312 & 17:57:28.26 & +55:23:12.08 & BL Lac & 14.8 & 0.065\\
		J180650+694928 & 18:06:50.68 & +69:49:28.09 & BL Lac & 11.1 & 0.046\\
		--- & --- & --- & --- & --- & ---\\
		\hline
	\multicolumn{6}{{|p{\columnwidth}|}}{\textbf{Note:} The entire table is available in online version. Only a portion of this table is shown here to display its form and content.}
	\end{tabular}
\end{table}

\begin{figure}
	\includegraphics[width=3.5in,height=3.5in]{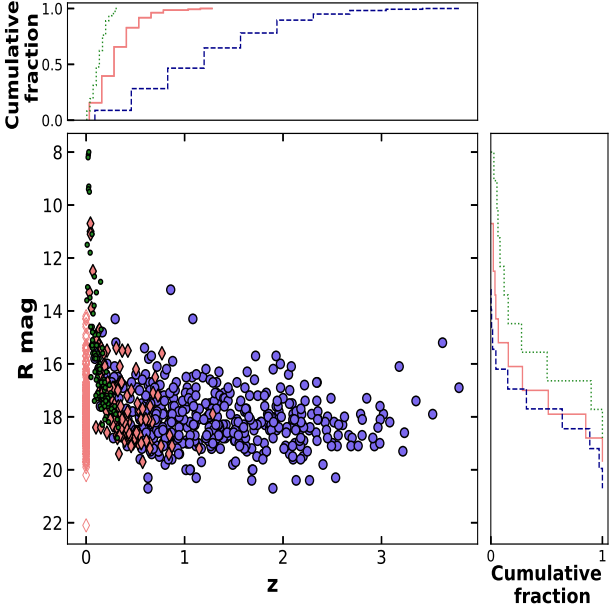}
    \caption{Distribution of the redshifts and magnitudes for the 897 blazars in our final sample. The targets with unknown redshift are plotted at $z=0$ with open faint red diamonds. The BL Lacs are denoted by red diamonds, the FSRQs by blue circles and the BL Lacs with host galaxy domination by green dots. The top and right-hand panels show the cumulative distributions of redshift and $R$-band magnitudes for the BL Lacs (solid red), FSRQs (dashed blue), and  BL Lacs with host galaxy domination (dotted green), respectively excluding the targets with unknown redshifts.}
    \label{fig:final_sample}
\end{figure}

The obtained light curves for all the blazars have a wide range of temporal spans ranging from $\sim$100 days to $\sim$1100 days with a typical time baseline of more than 700 days, or $\sim$2 years, for 94 per cent of the sources. Since blazars are highly variable sources with variability timescales from minutes to years, the lack of simultaneous multi-wavelength data is a severe obstacle in the broadband study of blazars \citep[e.g.,][]{2015RAA....15.1784Z}. To account for these effects, we have considered only the quasi-simultaneous observations to analyse the brightness variability and the true spectral behaviour of these sources. Therefore, we select the sources having at least 10 data points with quasi-simultaneous observations defined as being within 30 minutes in both $g$ and $r$ bands. This yields a final sample of 897 blazars, out of which 455 are BL Lacs, and 442 are FSRQs, with 109 of the BL Lacs having dominant host galaxy component in their SED. The redshift and $r$-band magnitude distributions of the final sample are shown in Fig. \ref{fig:final_sample}.

 \section{Analysis and Results}
\begin{figure*}
	\includegraphics[width=7.5in,height=5.5in]{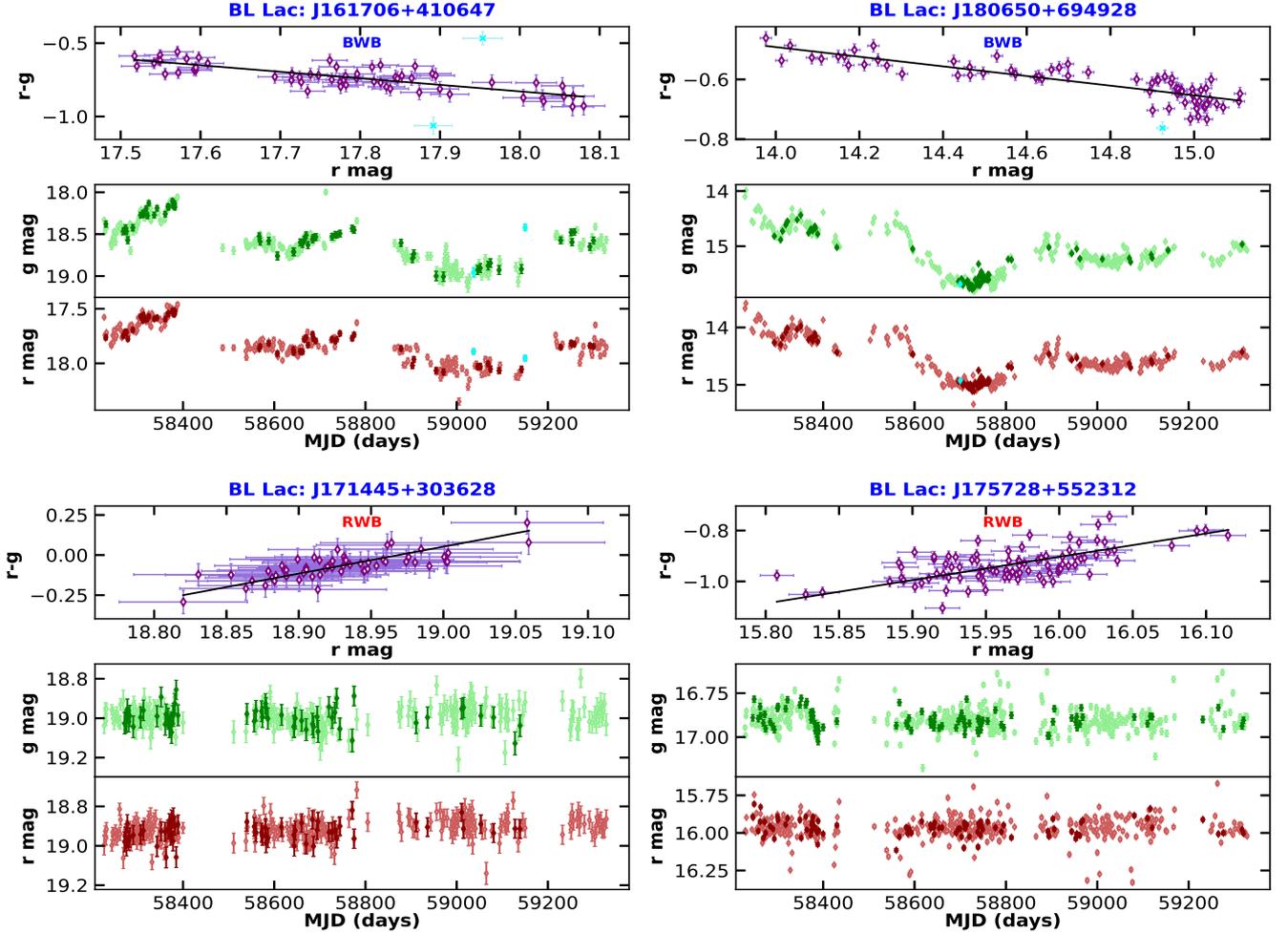}
    \caption{The $g$ and $r$ band ZTF light curves for 4 BL Lacs, (namely,  J161706+410647, J180650+694928, J171445+303628 and J175728+552312), from our sample, are shown in the middle and bottom panel of each subplot, respectively. The quasi-simultaneous observations used in the analysis are shown as filled diamonds, whereas the open diamonds represent the entire ZTF light curve.  The colour-magnitude diagrams are shown in the top panels. The solid line represents the best fit and the outliers (clipped points) are shown as cyan filled diamonds. The BL Lacs J161706+410647 and J180650+694928 follow a BWB trend over the complete duration of the observations, while the other two BL Lacs J171445+303628 and J175728+552312 follow an RWB trend.}
    \label{fig:all_bllac}
\end{figure*}

For each source, the observed data was first corrected for the foreground galactic interstellar reddening and absorption based on the extinction correction ($A_{v}$) provided by NASA's NED according to \citet{1998ApJ...500..525S}. The $g$ and $r$ band light curves for some of the BL Lacs and FSRQs from our sample are shown in the lower two panels of each subplot in Fig.~\ref{fig:all_bllac} and Fig.~\ref{fig:all_fsrq}.  The entire light curves from the ZTF survey, with MJD ranging from 58200 to 59450, are shown as open diamonds, whereas the quasi-simultaneous points used in our analysis are shown as filled diamonds. 
In what follows, we compare the brightness and colour variability of blazars using only the quasi-simultaneous observations in the $g$ and $r$ bands.

\subsection{Brightness Variability}
\label{sec:var_section}
 The diverse variability timescales in blazars may arise from various physical mechanisms, including accretion disc instabilities or the fluctuations arising in the jet. The flux variations in blazars can provide clues to the possible origin of variability. For instance, short term variability reported in flaring and high states are generally attributed to the shock moving down the inhomogeneous medium in the jet \citep{1985ApJ...298..114M,1995ARA&A..33..163W,2014ApJ...780...87M} whereas small amplitude variations seen in a low state of a blazar could be due to various instabilities in the accretion disc \citep[e.g.,][]{1991sepa.conf..557W,1993ApJ...411..602C,1993ApJ...406..420M}. Here, we first compare the brightness variations in the BL Lacs and FSRQs in our ZTF sample. To do so, we computed two frequently employed quantities: the amplitude of variability, $\Psi$, introduced by \cite{1996A&A...305...42H} and the fractional variability amplitude, $F_{var}$, described by \cite{2003MNRAS.345.1271V}. 
 First, we calculated the amplitude of temporal variability, $\Psi_{r}$,  using the Eq.~(\ref{eq:varamp_equation}) for the whole data set in the $r$ band.
 \begin{equation}
    \Psi_{r} = \sqrt{(A_{max}-A_{min})^{2}-2\sigma^{2}},
    \label{eq:varamp_equation}
\end{equation}
where $\sigma^{2} = < \sigma_{i}^{2}>$, with $A_{max}$ and $A_{min}$ being the maximum and minimum amplitudes in the light curve, and $\sigma_{i}$ being the uncertainty in the $i^{th}$ datapoint. \\   

\cite{2009ApJ...705...46B} have suggested imposing a variability cut at 0.4 mag to differentiate the stronger jet-based blazar fluctuations from quasar-like behaviour due to the accretion disc flares. It was found that 260 out of 455 BL Lacs in our sample, or $57.1$ per cent showed a variability amplitude $\geq$ 0.4 mag whereas only $34.6$ per cent (154/442) FSRQs showed a variability amplitude $\geq$ 0.4 mag. More extreme variability, with $\Psi_{r}$ $\geq$ 1 mag was shown by $17.4$ per cent (79) BL Lacs and $12.4$ per cent (55) FSRQs. Hence this analysis suggests that BL Lacs tend to be more variable, with the median variability amplitude of 0.47 in the $r$ band as compared to 0.30 in FSRQs. Some of the excess variability in BL Lacs, as compared to the FSRQs, is likely due to the larger observed contribution of the jet, which is closely aligned towards the observer and hence strongly Doppler boosted \citep{2020MNRAS.498.3578S};  the beaming amplifies the observed small-scale fluctuations and also reduces the timescale of these fluctuations. It may also be affected by the shorter observed rest-frame timescales for FSRQs in our sample as compared to BL Lacs as the former are distributed at higher redshifts, with 90 per cent of them having z $\geq$ 0.5, as compared to only 6 per cent in case of BL Lacs.\\\\
We also checked the fractional variability amplitude, $F_{var}$, of both types of blazars in the available $g$ and $r$ bands using the following Eq.~(\ref{eq:fvar_equation}). This gives a measure of intrinsic variability amplitude and represents the averaged amplitude of observed variations, corrected for the effects of measurement noise, as a fraction of the flux:
\begin{equation}
    F_{var} = \sqrt{\frac{S^{2} - \overline{\sigma_{err}^{2}}}{\overline{x^{2}}}}.
    \label{eq:fvar_equation}
\end{equation}
     Here the mean square error $\overline{\sigma_{err}^{2}}$ is given as: \\\\
    $\overline{\sigma_{err}^{2}} = \frac{1}{N} \sum\limits_{i=1}^{N} \sigma_{err,i}^{2}$
    \hspace{0.8cm} and \vspace{0.5cm} $S^{2} = \frac{1}{N-1} \sum\limits_{i=1}^{N}(x_{i}-\overline{x})^{2} $ is the sample variance, for the $N$ flux measurements, $x_i$.
    The error in $F_{var}$ is given by:
 \begin{equation}
 \sigma (F_{var}) = \sqrt{\left( \sqrt{\frac{1}{2N} }\frac{\overline{\sigma_{err}^{2}}}{\overline{x^{2}} F_{var}} \right)^{2} + \left( \sqrt{ \frac{\overline{\sigma_{err}^{2}}}{N} }\frac{1} {\overline{x}}  \right)^{2}} .
    \label{eq:fvarerr_equation}
\end{equation}

\begin{figure}
	\centering
    \includegraphics[width=\columnwidth]{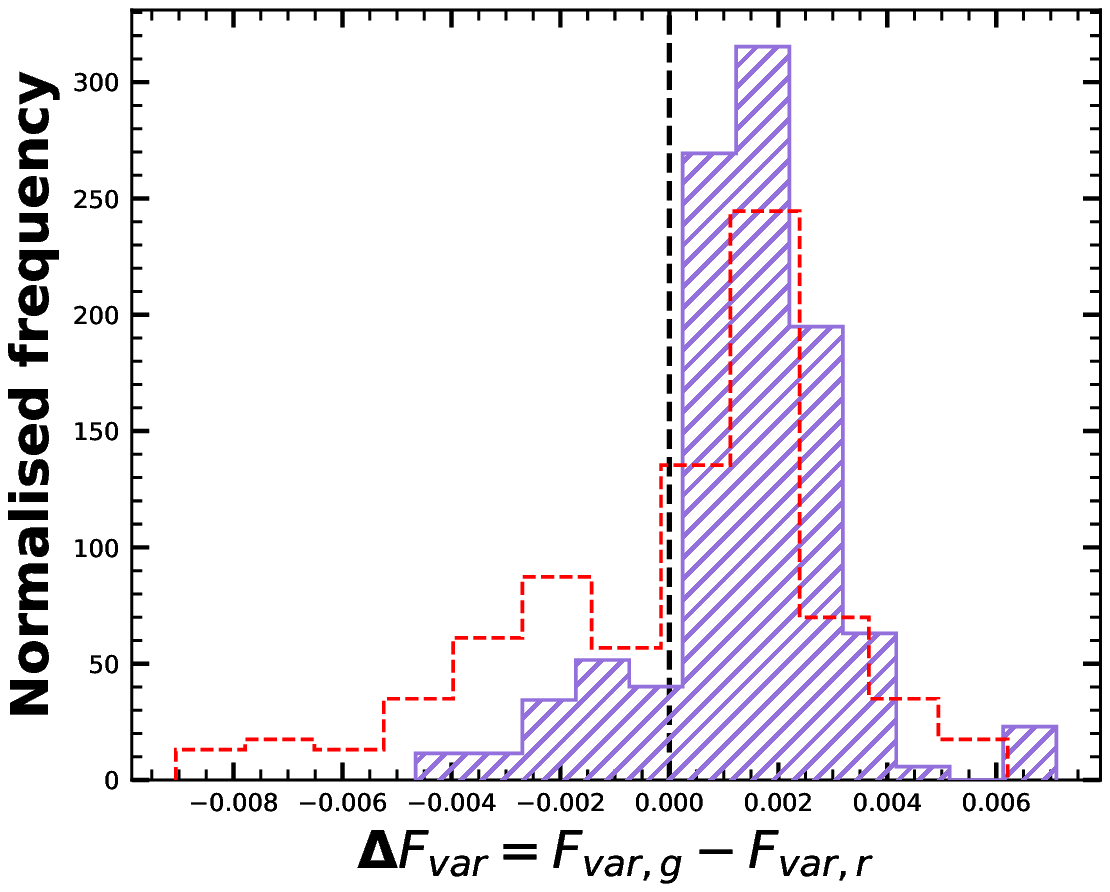}
  \caption{The differential fractional variability at the $g$ and $r$ bands ($\Delta F_{var}$) distributions for BL Lacs and FSRQs exhibiting $\Delta F_{var}$ at 3$\sigma$ level. The BL Lacs are shown in solid blue while the FSRQs are in dashed red. Each bin displays the bin's raw count divided by the total number of counts and the bin width.}
  \label{fig:quasi_Fvar_diff_distribution}
\end{figure}

 We find  358 ($\sim$40 per cent) of the blazars, consisting of 178 BL Lacs and 180 FSRQs, showed a significant fractional variability defined as exceeding the 3$\sigma $ level, whereas a  539 ($\sim$60 per cent), were found to have non-significant variability by this $F_{var}$ criterion during these ZTF observations.  
 Among the 178 BL Lacs, the $F_{var}$ in the $g$ band is found to be higher than that in the $r$ band for 152 objects and lower for only 26 objects. Similarly, out of the 180 FSRQs, 115 objects have higher $F_{var}$ in the $g$ band than in the $r$ band, whereas 65 sources have shown the opposite trend. The distributions of $\Delta F_{var}$ (i.e., $F_{var,g} - F_{var,r}$) for the BL Lacs and FSRQs are plotted in Fig. \ref{fig:quasi_Fvar_diff_distribution}. It can be seen that the $\ F_{var}$ in the higher frequency $g$ band is larger than that in the $r$ band for the majority of the blazars, with the fraction of FSRQs showing such a trend being relatively lower, i.e., $\sim$64 per cent, as compared to the BL Lacs (i.e., $\sim$85 per cent). 
 
\subsection{Colour Variability}
In addition to their rapid flux variations, blazars are also found to show spectral variations. The optical emission from blazars has contributions from the quasi-thermal emission from the disc and the non-thermal synchrotron emission from the relativistic jet.
Studying the spectral variations of blazars can help us in distinguishing the different components contributing to the observed flux. Since the colour variability of the blazars reflects their spectral variability,  the multi-band quasi-simultaneous observations of the large sample of blazars discussed here can help us in finding any universality in the spectral variations of the blazar population on substantial timescales of months to years. We note that if there are significant rapid variations over timescales of less than an hour, our requirement that the different band measurements be made within 30 minutes of each other would be insufficient, but only a small fraction of our data is so confounded.

\begin{figure*}
\includegraphics[width=7.5in,height=5.5in]{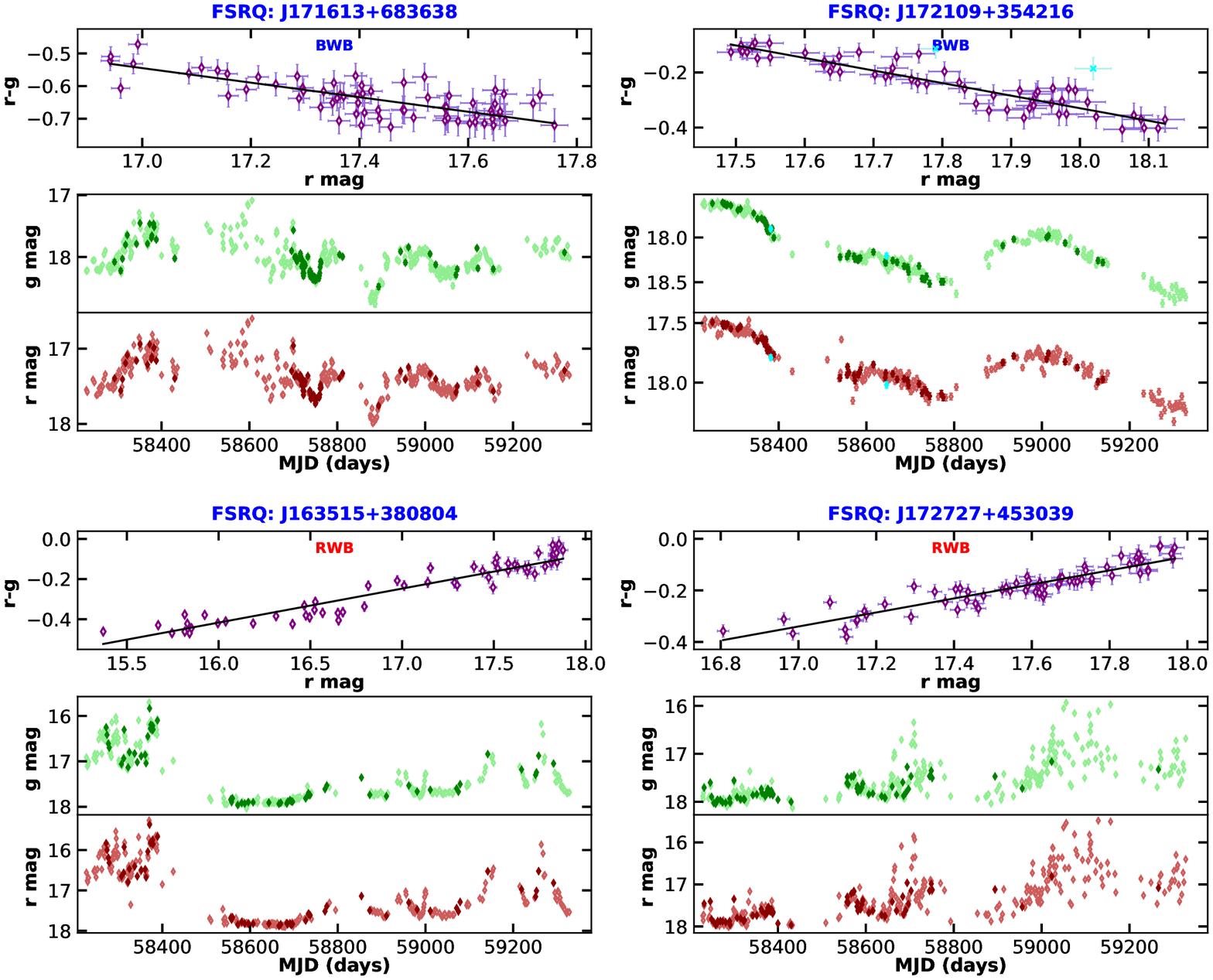}
    \caption{Same as Fig.~\ref{fig:all_bllac} but for four FSRQs (namely, J171613+683638, J172109+354216, J163515+380804 and J172727+453039) from our sample. The FSRQs J171613+683638 and J172109+354216 follow a BWB trend throughout the observations while the other two  follow a RWB trend.}
    \label{fig:all_fsrq}
\end{figure*}

For the 897 sources in our sample 
 we generated the colour-magnitude ($r$-$g$ vs $r$) diagrams using only the quasi-simultaneous data, as shown in top panels of each subplot of Fig.~\ref{fig:all_bllac} and Fig.~\ref{fig:all_fsrq}. The $g$ and $r$ band light curves are also shown for the respective sources in the middle and bottom panels of each subplot, where the quasi-simultaneous points are denoted by filled diamonds and the remainder of the light curve is shown in open diamonds. We carried out orthogonal distance regression fitting to the data points in the colour-magnitude diagram taking into account the uncertainties in both colour and magnitude; we also applied 3-sigma clipping on the residual points (i.e., model$-$observed) in order to remove the probable outliers in the colour-magnitude diagram.
 
 The best fit colour-magnitude models for a few of our sources are shown as solid straight lines in the top panels of Fig.~\ref{fig:all_bllac} and Fig.~\ref{fig:all_fsrq}, and the clipped points are shown in cyan cross symbols. In addition, the colour-magnitude diagrams for all the sources showing some trends in our sample are presented in the online material. For a robust measurement of possible bluer when brighter (BWB) or redder when brighter (RWB) trends in the light-curves, we first calculated the Pearson correlation coefficients by randomly redistributing the values of colour index (CI) and $r$ mag within the range from $P - P_{e}$ to $P + P_{e}$ for about $10^{4}$ realisation, where $P$ and $P_{e}$ are the parameter value and the corresponding error, respectively. The so obtained $10^{4}$ Pearson correlation coefficients were found to be well approximated by the Gaussian distributions, so the peak and the standard deviation of the modelled Gaussian function is considered as the final Pearson correlation coefficient and associated uncertainty, respectively. 

We have considered the colour behaviour of a source to be strong BWB if the Pearson correlation coefficient $\rho$ $\leq$ -0.5 and if it returns a rejection probability $p$ $\leq$0.05. In addition, a strong RWB trend is considered when $\rho$  $\geq$ 0.5 and $p$  $\leq$ 0.05. Using the above criteria, we found that $\sim$18.5 per cent (84/455) BL Lacs showed a strong BWB trend, whereas only $\sim$9.0 per cent (41/455) BL Lacs showed a RWB trend over the entire durations of these light curves. 
Interestingly, we found that for the BL Lacs showing a BWB trend, only $\sim$12 per cent (10/84) were contaminated by significant host galaxy contributions, whereas for the BL Lacs showing an RWB trend, about 71 per cent (29/41) were so contaminated. On the contrary, among a total of 442 FSRQs, a small fraction, i.e., $\sim$10.1 per cent (45/442) of systems showed a BWB trend, whereas $\sim$17.7 per cent (78/442) showed an RWB trend over the complete time duration available. Note that about 649 sources did not show any strong correlation between the CI and the $r$ band magnitude. However, if we relax our stringent criteria to allow for a Pearson correlation coefficient of $|{\rho}| \geq $ 0.2 and a null hypothesis rejection probability $p \leq 0.10$ respectively, the fractions of BL Lacs showing BWB and RWB trends increases to $\sim$29.7 per cent and $\sim$14.5 per cent, whereas for FSRQs it reaches $\sim$18.3 per cent and $\sim$26.7 per cent, respectively. \\

\begin{figure*}
	\includegraphics[width=7.5in,height=5.5in]{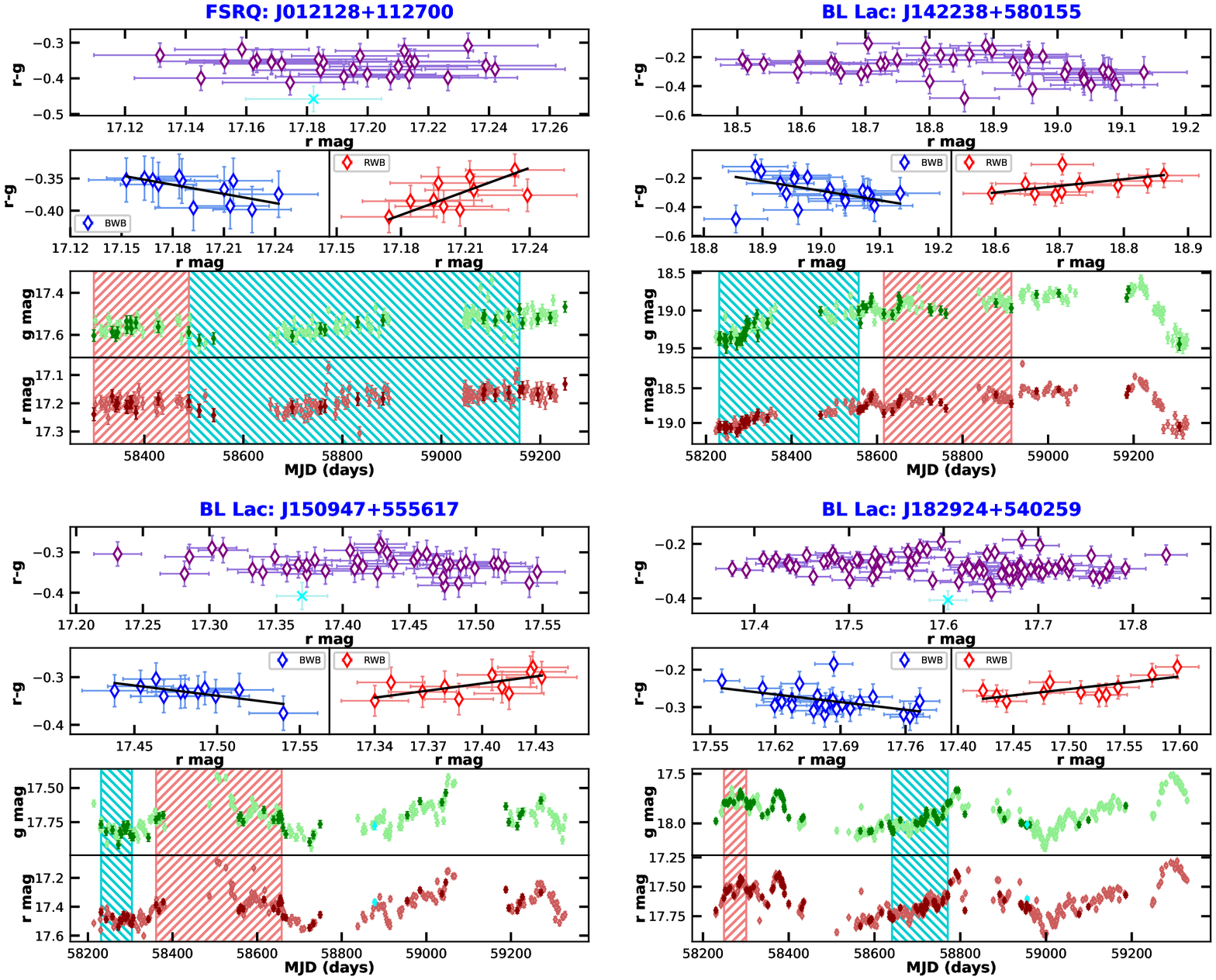}
    \caption{The partial multi-colour variation trends observed in the blazars J012128+112700, J142238+580155, J150947+555617 and J182924+540259. The $g$ and $r$ band light curves are shown in the bottom two panels of each subplot, respectively. The intervals with BWB and RWB trends are shown as a slanted region at $-$45 degrees and 45 degrees, respectively. The colour-magnitude diagrams for the partial trends are shown in the second panel from the top, along with the best fit (solid line). The colour-magnitude diagram for the entire temporal span is shown in the top panel.}
    \label{fig:all_partial}
\end{figure*}

For a handful of blazars, a partial trend in colour behaviour has been reported \citep{2011PASJ...63..639I, 2017ApJ...844..107I}. These blazars have shown either a trend for only a short duration or different trends at different epochs. These different behaviours at different times can possibly occur as a result of a change in the relative intensities of the accretion disc and jet emissions. Given the good quality of the available ZTF data for such analyses, we have also searched for the partial trends in the colour behaviour for the entire sample.
To do so, we investigated the colour-magnitude correlation over different timescales by binning the light curve and ensuring that at least 10 continuous quasi-simultaneous data points contribute to the correlation. The partial trend is calculated at all possible timescales shorter than the available $\sim$2 years duration but exceeding at least 10 days with one point observed each night, keeping the stringent correlation criteria (with  $\rho \geq 0.5$ or $ \rho \leq -0.5$ and $p \leq 0.05$).

With this approach, out of the 330 ($\sim$72.5 per cent) of BL Lacs showing no colour trends over the entire light curves, 54 showed a partial BWB trend at some interval of time, whereas 72 showed a partial RWB trend.  Additionally, 8 BL Lacs showed a complex colour behaviour consisting of both partial BWB and RWB trends for different lengths of time, as shown in the second panel from the top for each subplot in Fig.~\ref{fig:all_partial}. The duration over which the BWB or RWB trends were observed are shown as slanted lines at 45 degrees and $-$45 degrees, respectively, in the $g$ and $r$ band light curves (see, Fig.~\ref{fig:all_partial}, lower two panels in each subplot). In the case of FSRQs, out of the 319 sources showing no trend across the whole duration, 29 showed a partial BWB trend, 120 showed a partial RWB trend, and 8 showed a complex behaviour with both trends at different intervals of time. Full details about the trends followed by all the sources in our sample are given in Table~\ref{tab:result_table} along with their r-band variability amplitude, $\Psi_{r}$ and the differential fractional variability, $\Delta F_{var}$.
\begin{table}
	\centering
	\caption{The number of BL Lacs and FSRQs showing BWB or RWB trends in the complete data set. The numbers in brackets show the BL Lacs dominated by host galaxies. A strong correlation is defined as Pearson correlation coefficient $ \rho  \geq 0.5$ or $ \rho  \leq -0.5$ and $\geq 0.95$ confidence whereas a weak correlation is at $\rho  \geq 0.2$ or $\rho  \leq -0.2$ and $\geq 0.90$ confidence for the typically $\sim$2 years lengths of the observations. The partial trends require   $\rho  \geq 0.5$ or $ \rho  \leq -0.5$ and $\geq 0.95$ confidence for durations exceeding 10 days but shorter than the full observations.}
	\label{tab:final_table}
	\begin{tabular}{|c | c |c |c |c |c |c|} 
		\hline
		Correlation & \multicolumn{2}{c}{Strong} & \multicolumn{2}{c}{Weak} &  \multicolumn{2}{c}{Partial}\\
		\cline{1	$\-$ -1}
		\cline{2	$\-$ -3}
		\cline{4	$\-$ -5}
		\cline{6	$\-$ -7}
		Trend & BWB & RWB & BWB & RWB & BWB & RWB\\
		\hline
		BL Lac & 84(10) & 41(29) & 135(11) & 66(46) & 54(1) & 72(21)\\
		FSRQ & 45 & 78 & 81 & 118 & 29 & 120\\
		\hline
	\end{tabular}
\end{table}

\begin{figure*}
	\includegraphics[width=5.5in,height=4.5in]{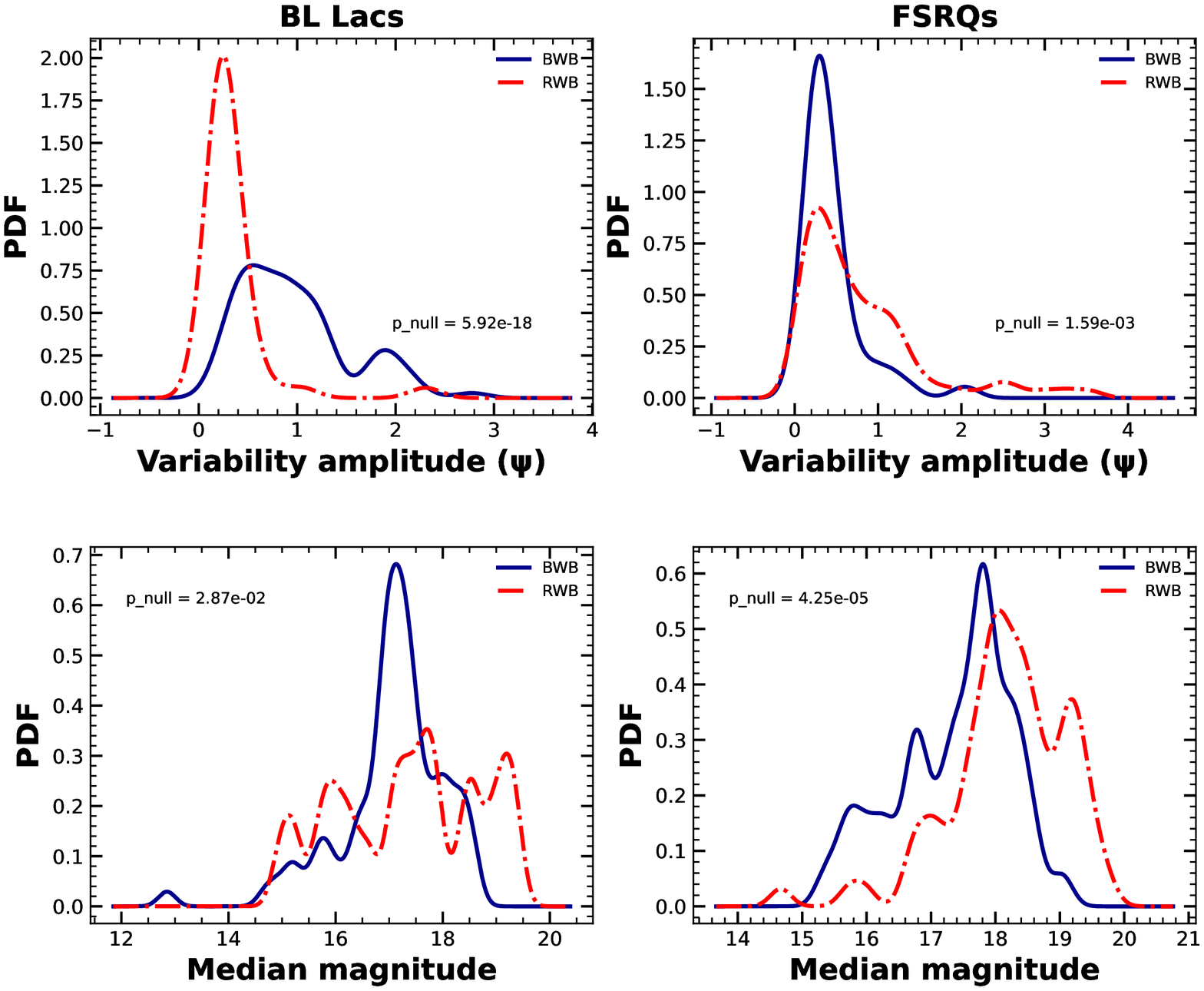}
   \caption{
Distributions of  Gaussian kernels are shown for the variability amplitude in the top row and $r$-band median magnitudes in the bottom rows of panels for the BL Lacs following a BWB (solid line) and RWB (dotted line) in the left column and for FSRQs in the right column of panels. The probability distribution function (PDF) is derived using a non-parametric kernel density estimation (KDE) \citep[see][]{1986desd.book.....S} with a fixed bandwidth derived using maximum likelihood cross-validation approach for each distribution. The $p_{null}$ computed using the K-S test between the BWB and RWB subsamples are given in bottom right/ top left corners. }
\label{fig:Fig_7_8}
\end{figure*}

\subsection{Correlation of colour behaviour with other parameters}
To try to understand the origin of colour behaviour in blazars, we compare the $r$-band variability amplitude for all the BL Lacs and FSRQs showing different colour behaviours for the entire duration of these observations. The distributions of variability amplitudes are shown in the top panels of Fig.~\ref{fig:Fig_7_8}. We see that the BL Lacs showing BWB behaviour tend to show higher variability amplitudes as compared to those exhibiting RWB trends, with median $r$-band variability amplitudes of $\sim$0.87 mag and $\sim$0.25 mag, respectively. We also performed the two-sample Kolmogorov-Smirnov (KS) test on the variability amplitudes of these two samples and found the $p_{null}$ of $5.9 \times 10^{-18}$, implying that the two samples are derived from different populations. The corresponding median variability amplitudes for the distributions of FSRQs showing BWB and RWB are found to be $\sim$0.31 mag and $\sim$0.55 mag with $p_{null}$ of only $1.5 \times 10^{-3}$. 

We also compare the $r$-band median magnitude distributions of these subsamples in the bottom panels of Fig.~\ref{fig:Fig_7_8}. We find that the median magnitude for FSRQs with BWB trends is smaller than the ones following RWB trends, with values of 17.36 and 18.17, respectively. The KS test on the distribution of median magnitudes of BWB and RWB FSRQs, results in the $p_{null}$ of $4 \times 10^{-5}$ indicating that the two samples are different. This hints that the brighter FSRQs, likely to be in more active states, follow a BWB trend, whereas the fainter ones follow an RWB trend.

\begin{figure*}
	\includegraphics[width=5.55in,height=2.50in]{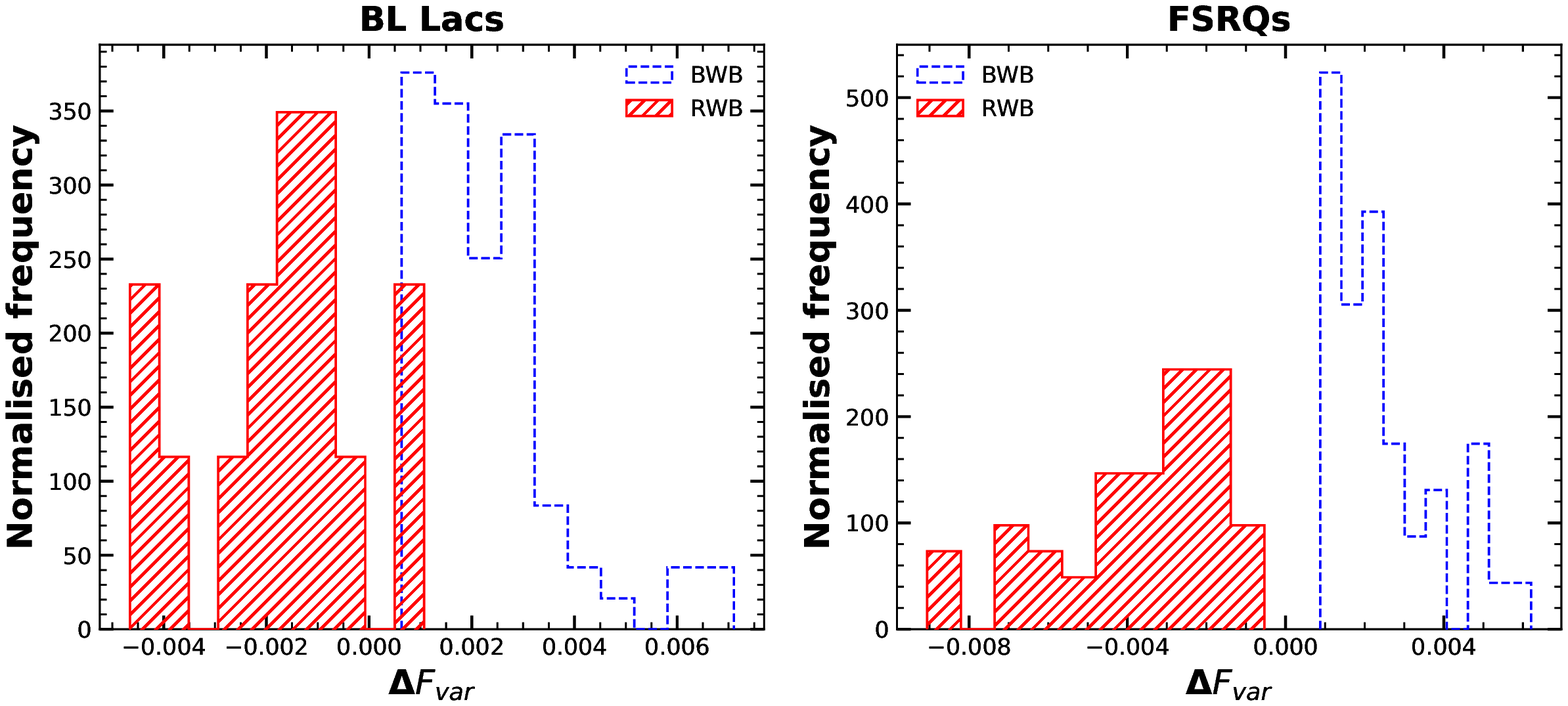}
 \caption{Left Panel: Distribution of $\Delta F_{var}$ for the BL Lacs following the BWB (dashed blue line) and RWB (solid red line) trends. Each bin displays the bin's raw count divided by the total number of counts and the bin width. Right Panel: The same for FSRQs.}
 
 \label{fig:delta_Fvar_BWB_RWB}
\end{figure*}

We also  examined the difference between the $g$ and $r$ bands fractional variability amplitudes defined as $\Delta F_{var} =  F_{var,g} - F_{var,r}$ for the sources showing BWB and RWB trends in the $\sim$2 years timescale. Fig.~\ref{fig:delta_Fvar_BWB_RWB} shows the $\Delta F_{var}$ distribution for both BWB and RWB BL Lacs and FSRQs. It was found that all the BL Lacs and FSRQs showing a BWB trend have $\Delta F_{var}$ > 0, whereas all sources showing RWB trend have $\Delta F_{var}$ < 0. This demonstrates in another way that differences in variability amplitudes lead to different colour behaviours. In other words, for the sources with a simultaneous flux variations, a higher amplitude of variability at a higher frequency band goes along with the BWB trend, and those having larger amplitude variations at a lower frequency can give rise to the RWB trend.

\section{Discussion}

Optical flux and colour variability long have been a subject of blazar research, with many attempts made to study the origin of their flux variations. Multiple studies have investigated the colour behaviour of the blazars using different blazar samples and individual blazars over different timescales \citep[e.g.,][]{2003ApJ...590..123V,2006A&A...450...39G,2010MNRAS.404.1992R,2011PASJ...63..639I,2016Ap&SS.361..345M}.
However, due to the lack of reasonably unbiased and large samples, the universality in their colour behaviour has been highly debated.
In this paper, we have been able to search for any commonalities in the colour behaviour of blazars by examining the largest sample of these AGN for this purpose to date. 

Investigating the colour behaviour of blazars on the rather large timescale of $\sim$2 years, we found that for the BL Lacs, where the jet is likely aligned along the line of sight within $10^{\circ}$ \citep{1995PASP..107..803U}, 84 (out of 455 BL Lacs) targets show the BWB trend. This is about a factor of 2 higher than the 41 BL Lacs showing the RWB trend. We note that out of these 41  RWB trending BL Lacs, 29 of them or $\sim$71 per cent, were found to be contaminated by their host galaxy emission. Further, the subsample of BL Lacs showing the BWB trend are found to exhibit higher variability amplitude in comparison to the subsample of BL Lacs showing the RWB trend. On the contrary, a relatively larger number of FSRQs, 78 out of 442, showed a RWB trend   than a BWB trend (45 FSRQs). We also found that the FSRQs showing BWB trend have median magnitudes relatively brighter than the RWB ones which hints that, similarly to BL Lacs, the FSRQs also tend to follow a BWB trend in their brighter phases.  Moreover, the majority of our BL Lacs showing RWB trends were found to have large host galaxy contaminations. As most blazars are hosted by elliptical galaxies \citep[e.g.,][]{2000ApJ...544..258S} that are mainly composed of red giants and stars on the asymptotic branch and have a very small fraction of hot young blue stars, the host galaxies emission components are redder. 
We note that for extended sources, such as nearby AGNs, a change in seeing may affect the host galaxy contribution \cite[e.g.][]{2017ApJ...849..161F}. However, the typical seeing at the epoch of observations for the host galaxy dominated sources  in our sample remained between 1.5 arcsec and 2.5 arcsec for the majority of the time and their color indices do not show any noticeable dependence on this nominal seeing variation that could lead to any significant impact on the observed RWB/BWB trends.

Previous colour variability studies based on optical bands of a handful of blazars have reported different colour behaviours on different timescales. \cite{2003ApJ...590..123V} found that eight BL Lacs in their sample exhibited the BWB trend in the optical bands. \cite{2006A&A...450...39G} monitored 8 red blazars from Sep 2003 to Feb 2004 and found that 5 BL Lacs showed the BWB trend, 2 FSRQs displayed the RWB trend and 1 FSRQ was achromatic over a period of five months. In a 10 month-long monitoring campaign of twelve blazars, \cite{2010MNRAS.404.1992R} found that three out of six BL Lacs exhibited the BWB trend, whereas four out of six FSRQs showed the opposite RWB trend. These studies led to an interpretation that the BWB trend is generally associated with BL Lacs whereas, in FSRQs, an RWB trend is the norm. Our analyses of the colour behaviour of a much larger sample of blazars over longer durations have also shown that the BWB trend is more prominent in the case of BL Lacs whereas, for FSRQs, the RWB trend is more frequently observed. The predominance of the BWB trend seen in the BL Lacs can be explained by the shock-in-jet model where the electrons accelerated to higher energies at the front of a shock lose energy faster due to radiative cooling and thus make the high energy bands more variable \citep{1998A&A...333..452K,2002PASA...19..138M}. It can also be explained if the luminosity increase was due to the injection of fresh electrons with an energy distribution harder than that of the previous, partially cooled ones.
FSRQs are low synchrotron peaked blazars with peak synchrotron frequencies lying in the infrared regime, whereas their accretion disc components peak in the optical to ultraviolet \citep{2017A&ARv..25....2P}. Therefore the usual RWB trend observed in the FSRQs is likely because of the addition of more variable and redder non-thermal jet emission to an already blue, quasi-thermal disc component {\citep{2006A&A...450...39G,2010MNRAS.404.1992R}}.

\cite{2011PASJ...63..639I} studied 42 blazars consisting of 29 BL Lacs and 13 FSRQs for a timescale of $\sim$2 years and detected a BWB trend in a high fraction, 66\% (i.e., 28/42) of the blazars and RWB trend in only 3 of their objects. For these three objects, they also found a BWB trend in their brighter phases. They proposed that the BWB trend is universal in blazars, and two factors, the presence of a thermal disc and multiple variable synchrotron components can disturb the general BWB trend in the blazars.
The FSRQs in our sample also show similar trends between their magnitudes and their colour behaviour, with the brighter FSRQs often following a BWB trend, whereas the fainter ones much more frequently exhibit an RWB trend, as shown in Fig.~\ref{fig:Fig_7_8}. This can occur if the FSRQs showing RWB trends have been observed in their faint state and contain a significant contribution of thermal emission from the disc, whereas FSRQs showing BWB trends have been observed in the bright state when they are completely dominated by non-thermal jet emission, as are BL Lacs.
Another study by \cite{2016Ap&SS.361..345M} on 29 Blazars, consisting of 25 BL Lacs and only 4 FSRQs found that 17 out of 25 BL Lacs and 2 out of 4 FSRQs showed a significant BWB trend in their stripe 82 observational periods, whereas 1 FSRQ followed an RWB trend. They propose that the BWB trend is quite common in long-term optical light curves of BL Lacs while an RWB trend in FSRQs is attributed to the thermal emission from the accretion disc. The framework presented in their study is also consistent with our analysis that the different relative contributions of the thermal versus non-thermal radiation to the total optical emission may be responsible for the different trends of the CI with brightness in BL Lacs and FSRQs.

\begin{table*}
	\centering
	\caption{Retrieved parameters and the trends for all the sources in our sample. }
	\label{tab:result_table}
	\begin{tabular}{cccccccc}
		\hline
		Object ID & \multicolumn{1}{c}{Name} & Type &   Trend & Duration & Variability amplitude  &   \multicolumn{2}{c}{Differential fractional variability}\\ 
		          &                          &      &         &           & \multicolumn{1}{c}{$\Psi_{r}$}& \qquad $\Delta F_{var}$    & $\sigma(\Delta F_{var})$\\		
		\hline

        1 &	J225005+382437 &	BL Lac &	none    &   ---   &  \multicolumn{1}{c}{0.28}   &	 \qquad 0.000973   &	0.000238\\
        2 &	J220243+421640 &	BL Lac &	BWB    &   entire    &  \multicolumn{1}{c}{0.26}   &	 \qquad 0.001730   &	0.000576\\
        3 &	J132802+112913 &	BL Lac &	none    &   ---   &  \multicolumn{1}{c}{0.46}   &	 \qquad 0.001499   &	0.000487\\
        4 &	J151041+333504 &	BL Lac &	RWB    &   entire    &  \multicolumn{1}{c}{0.19}   &	 \qquad -0.001892   &	0.000671\\
        5 &	J041652+010523 &	BL Lac &	none    &   ---  &  \multicolumn{1}{c}{0.54}   &	 \qquad -0.000515   &	0.000614\\
        6 &	J232012+414605 &	BL Lac &	RWB    &   partial  &  \multicolumn{1}{c}{0.29}   &	 \qquad 0.000094   &	0.000463\\
        7 &	J083133+174630 &	BL Lac &	none    &   ---   &  \multicolumn{1}{c}{0.34}   &	 \qquad 0.001061   &	0.000564\\
        8 &	J181050+160820 &	BL Lac &	none    &   ---   &  \multicolumn{1}{c}{1.53}   &	 \qquad -0.000024   &	0.000509\\
        9 &	J144248+120040 &	BL Lac &	BWB    &   entire    &  \multicolumn{1}{c}{0.87}   &	 \qquad 0.003729   &	0.000219\\
        10 &	J083223+491321 &	BL Lac &	BWB    &   entire    &  \multicolumn{1}{c}{1.24}   &	 \qquad 0.003111   &	0.000348\\
        

		--- & \multicolumn{1}{c}{---} & --- & --- & --- & \multicolumn{1}{c}{---} & \qquad --- & ---\\
		\hline
	\multicolumn{8}{{|p{1.8\columnwidth}|}}{\textbf{Note:} The entire table is available in online version. Only a portion of this table is shown here to display its form and content.}
	\end{tabular}
\end{table*}

Interestingly, we have also found that the BL Lacs showing BWB trends are significantly more variable than the BL Lacs showing RWB trends as shown in Fig.~\ref{fig:Fig_7_8}, whereas, in the case of FSRQs, we found no such significant correlation.
 A very long-term study ($\sim$10 years) of 37 blazars by \cite{2004A&A...419...25F} in the optical bands with quasi-simultaneous observations ($\Delta$t < 2hrs) have also found that the bluer-when-brighter trend is more prominent in extremely variable objects. However, they also mentioned that their sample was biased toward BL Lacs, with only 6 FSRQs in the sample. Hence, we propose that the most extremely variable BL Lacs are more prone to show the BWB trend, but no such claim could be made for FSRQs. This goes in accordance with the proposition that the BWB trends could be due to the presence of two components contributing to the overall emission in the blazar's optical flux, one variable (with a flatter slope) and the other stable (with $\alpha_{stable}$ > $\alpha_{var}$). However, this result can also be understood with a one-component synchrotron model framework: the more intense the energy release, the higher the particle's energy, as proposed in \cite{2004A&A...419...25F}. 

We also found that partial colour variations, i,e., a BWB trend at certain epochs and an RWB trend at other epochs or achromatic at certain epochs and chromatic at other epochs, is a  common phenomenon in blazars; previously, it had been seen in only a handful of blazars  \citep{2017ApJ...844..107I,2019MNRAS.486.1781R}. For instance, \cite{2017ApJ...844..107I} found that the optical-infrared colour variability of FSRQ 3C 279 altered between RWB, BWB, and achromatic on different short-timescales. From our sample of 897 blazars, we found 8 BL Lacs and 8 FSRQs showing multiple colour behaviours at different epochs. Additionally, 54 BL Lacs and 29 FSRQs showed a partial BWB trend and 72 BL Lacs and 120 FSRQs showed a partial RWB trend for a smaller duration but were achromatic in the longer duration of $\sim$2 years. This fluctuation in colour behaviour is likely due to the continuous variation in the contribution of the thermal accretion disc and non-thermal relativistic jet. The prevalent colour variation in blazars can also be understood within a general framework proposed by \cite{2017ApJ...844..107I}, where transitions from disc dominated, to a mixed contribution, to a jet dominated system, can drive the shift in the colour behaviours of the source at different times. Another possible explanation for the sources being achromatic on the longer duration but showing partial trends is if modest colour variations present during shorter periods are smeared out over the larger time window of the complete dataset. For such cases, distinguishing separate states in the colour-magnitude diagram can help in detecting any such trend in their colour behaviour.

It is worth noting that the fraction of BL Lacs showing the BWB and RWB trends increase by 11.2 per cent and 5.5 per cent, respectively after relaxing the Pearson correlation coefficient threshold from $\pm$0.5 to $\pm$0.2 and the rejection probability from $\pm$0.05 to $\pm$0.10. Similarly, the fraction of FSRQs showing BWB and RWB trends increase by 8.3 per cent and 7 per cent, respectively. The observed weak correlations in these systems may be due to the small wavelength difference between the $g$ and $r$ bands. For a portion of the blazars in our sample that show no correlation, it is possible that the source is in the state where jet and disc contributions are comparable,  as suggested by \citet{2017ApJ...844..107I} and therefore, there is some cancellation of the trends observed when one or the other is dominant.  

\section{Conclusions}

We have searched for optical flux and spectral variability in a large sample of blazars, including 455 BL Lacs and 442 FSRQs using observations made with the  ZTF  quasi-simultaneously (within 30 minutes) in $g$ and $r$ bands. We demanded at least 10 such quasi-simultaneous measurements for each source to include it in our sample, though the numbers of them are usually much greater. For most sources, the observations were taken over time spans of $\sim$2 years and we also looked at trends on shorter timescales. The key conclusions of our analysis are as follows:
\begin{itemize}
    \item A significant variability is found in both subclasses of blazars, with BL Lacs showing higher $r-$band median variability amplitude than FSRQs with $\Psi_{r}$ = 0.47 and 0.30 mag, respectively, implying that the physical conditions in these two populations are different. Moreover, the fraction of BL Lacs showing extreme variability (i.e., $\Psi_{r}$ > 1 mag) is relatively greater (0.17) than for FSRQs (0.12). The excess $r$-band variability in BL Lacs with respect to FSRQs is likely driven by the strong observed jet contribution, arising from the close alignment of the jet towards the observer.
   
    \item A significant fractional variability amplitude, with significance defined as $F_{var}$ exceeding a 3$\sigma$ level, is seen in only 40 per cent of blazars in this sample. Among them, we find a  systematically larger $F_{var}$ at $g$ band for $\sim$85 per cent of the BL Lacs and $\sim$64 per cent  of FSRQs, indicating that optical variability is frequency-dependent. However, this was not the case for all the sources, as only 152 out of 455 BL Lacs and 115 out of 442 FSRQs showed a larger $F_{var}$ at $g$ band than at $r$ band, whereas 26 BL Lacs and 65 FSRQs showed a higher $F_{var}$ at $r$ band than that at $g$ band.
   
    \item We found that over the longest probed timescale of about 2 years, among 455 BL Lacs, 84 systems showed a BWB trend. For the 41 BL Lacs showing RWB trend, about 71 per cent are found to be host galaxy dominated. This indicates that the non-thermal jet synchrotron emission dominates the optical continuum in BL Lacs and since the emission from the host galaxy peaks in the redder bands, this disturbs the BWB trend of these BL Lacs. Moreover, on correlating the colour behaviour of BL Lacs with the variability amplitudes in the $r$ band, we found that the extremely variable BL Lacs are most likely to show the BWB trend.

    \item For the 442 FSRQs, only 45 systems showed a BWB trend, whereas 78 evinced a RWB trend, indicating that the optical emission in FSRQs includes strong contributions from the quasi-thermal emission from the accretion disc. Additionally, the RWB FSRQs are found to be fainter in magnitudes as compared to the BWB FSRQs, implying that FSRQs in their bright states have greater contributions from the jet, whereas in their fainter states, the contribution from the accretion disc increases such that they more likely follow a BWB trend in the bright state and RWB trend in the faint state.

    \item   A good fraction of blazars are found to show a trend over limited temporal spans. Specifically,  54 BL Lacs and 29 FSRQs illustrated partial BWB trends, whereas 72 BL Lacs and 120 FSRQs have shown partial RWB trends. In addition to those, 8 BL Lacs and 8 FSRQs have shown complex colour behaviours with both kinds of trends observed at different epochs. These types of changes could arise from transitions between jet-dominated and disc-dominated states. Detection of partial trends in such a high fraction of blazars suggests that shorter-lived trends are common when long-term colour behaviours of blazars are studied.

\end{itemize}

\section*{Acknowledgements}

The authors thank the anonymous referee for providing  useful comments and suggestions that led to the improvement of the quality of the paper. LCH was supported by the National Science Foundation of China (11721303, 11991052) and the National Key R\&D Program of China (2016YFA0400702). VN acknowledges discussions with Sapna Mishra and Jaydeep Singh on various technical aspects and also gratefully acknowledges the Department of Physics and Astronomical Sciences, CUHP, Dharamshala, for the hospitality. 

This work is based on observations obtained with the Samuel Oschin Telescope 48-inch and the 60-inch Telescope at the Palomar Observatory as part of the Zwicky Transient Facility project. ZTF is supported by the National Science Foundation under Grant No. AST-2034437 and a collaboration including Caltech, IPAC, the Weizmann Institute for Science, the Oskar Klein Center at Stockholm University, the University of Maryland, Deutsches Elektronen-Synchrotron and Humboldt University, the TANGO Consortium of Taiwan, the University of Wisconsin at Milwaukee, Trinity College Dublin, Lawrence Livermore National Laboratories, and IN2P3, France. Operations are conducted by COO, IPAC, and UW.

This research has made use of the NASA/IPAC Extragalactic Database (NED) which is operated by the Jet Propulsion Laboratory, California Institute of Technology, under contract with the National Aeronautics and Space Administration.

\section*{Data Availability}
The data used in this study is publicly available in the ZTF DR6.


\bibliographystyle{mnras}
\bibliography{references} 

\begin{thebibliography}{}
\makeatletter
\relax
\def\mn@urlcharsother{\let\do\@makeother \do\$\do\&\do\#\do\^\do\_\do\%\do\~}
\def\mn@doi{\begingroup\mn@urlcharsother \@ifnextchar [ {\mn@doi@}
  {\mn@doi@[]}}
\def\mn@doi@[#1]#2{\def\@tempa{#1}\ifx\@tempa\@empty \href
  {http://dx.doi.org/#2} {doi:#2}\else \href {http://dx.doi.org/#2} {#1}\fi
  \endgroup}
\def\mn@eprint#1#2{\mn@eprint@#1:#2::\@nil}
\def\mn@eprint@arXiv#1{\href {http://arxiv.org/abs/#1} {{\tt arXiv:#1}}}
\def\mn@eprint@dblp#1{\href {http://dblp.uni-trier.de/rec/bibtex/#1.xml}
  {dblp:#1}}
\def\mn@eprint@#1:#2:#3:#4\@nil{\def\@tempa {#1}\def\@tempb {#2}\def\@tempc
  {#3}\ifx \@tempc \@empty \let \@tempc \@tempb \let \@tempb \@tempa \fi \ifx
  \@tempb \@empty \def\@tempb {arXiv}\fi \@ifundefined
  {mn@eprint@\@tempb}{\@tempb:\@tempc}{\expandafter \expandafter \csname
  mn@eprint@\@tempb\endcsname \expandafter{\@tempc}}}

\bibitem[\protect\citeauthoryear{{Agarwal} et~al.,}{{Agarwal}
  et~al.}{2019}]{2019MNRAS.488.4093A}
{Agarwal} A.,  et~al., 2019, \mn@doi [\mnras] {10.1093/mnras/stz1981}, \href
  {https://ui.adsabs.harvard.edu/abs/2019MNRAS.488.4093A} {488, 4093}

\bibitem[\protect\citeauthoryear{{Angel} \& {Stockman}}{{Angel} \&
  {Stockman}}{1980}]{1980ARA&A..18..321A}
{Angel} J.~R.~P.,  {Stockman} H.~S.,  1980, \mn@doi [\araa]
  {10.1146/annurev.aa.18.090180.001541}, \href
  {https://ui.adsabs.harvard.edu/abs/1980ARA&A..18..321A} {18, 321}

\bibitem[\protect\citeauthoryear{{Bauer} et~al.,}{{Bauer}
  et~al.}{2009}]{2009ApJ...705...46B}
{Bauer} A.,  et~al., 2009, \mn@doi [\apj] {10.1088/0004-637X/705/1/46}, \href
  {https://ui.adsabs.harvard.edu/abs/2009ApJ...705...46B} {705, 46}

\bibitem[\protect\citeauthoryear{{Bellm} et~al.,}{{Bellm}
  et~al.}{2019}]{2019PASP..131a8002B}
{Bellm} E.~C.,  et~al., 2019, \mn@doi [\pasp] {10.1088/1538-3873/aaecbe}, \href
  {https://ui.adsabs.harvard.edu/abs/2019PASP..131a8002B} {131, 018002}

\bibitem[\protect\citeauthoryear{{Blandford} \& {Rees}}{{Blandford} \&
  {Rees}}{1978}]{1978bllo.conf..328B}
{Blandford} R.~D.,  {Rees} M.~J.,  1978, in {Wolfe} A.~M.,  ed., BL Lac
  Objects. pp 328--341

\bibitem[\protect\citeauthoryear{{Bonning} et~al.,}{{Bonning}
  et~al.}{2012}]{2012ApJ...756...13B}
{Bonning} E.,  et~al., 2012, \mn@doi [\apj] {10.1088/0004-637X/756/1/13}, \href
  {https://ui.adsabs.harvard.edu/abs/2012ApJ...756...13B} {756, 13}

\bibitem[\protect\citeauthoryear{{Chakrabarti} \& {Wiita}}{{Chakrabarti} \&
  {Wiita}}{1993}]{1993ApJ...411..602C}
{Chakrabarti} S.~K.,  {Wiita} P.~J.,  1993, \mn@doi [\apj] {10.1086/172862},
  \href {https://ui.adsabs.harvard.edu/abs/1993ApJ...411..602C} {411, 602}

\bibitem[\protect\citeauthoryear{{Ciprini}, {Tosti}, {Raiteri}, {Villata},
  {Ibrahimov}, {Nucciarelli}  \& {Lanteri}}{{Ciprini}
  et~al.}{2003}]{2003A&A...400..487C}
{Ciprini} S.,  {Tosti} G.,  {Raiteri} C.~M.,  {Villata} M.,  {Ibrahimov} M.~A.,
   {Nucciarelli} G.,   {Lanteri} L.,  2003, \mn@doi [\aap]
  {10.1051/0004-6361:20030045}, \href
  {https://ui.adsabs.harvard.edu/abs/2003A&A...400..487C} {400, 487}

\bibitem[\protect\citeauthoryear{{Dermer}, {Schlickeiser}  \&
  {Mastichiadis}}{{Dermer} et~al.}{1992}]{1992A&A...256L..27D}
{Dermer} C.~D.,  {Schlickeiser} R.,   {Mastichiadis} A.,  1992, \aap, \href
  {https://ui.adsabs.harvard.edu/abs/1992A&A...256L..27D} {256, L27}

\bibitem[\protect\citeauthoryear{{Feng}, {Liu}, {Fan}, {Zhao}, {Bai}, {Wang},
  {Xiong}  \& {Li}}{{Feng} et~al.}{2017}]{2017ApJ...849..161F}
{Feng} H.-C.,  {Liu} H.~T.,  {Fan} X.~L.,  {Zhao} Y.,  {Bai} J.~M.,  {Wang} F.,
   {Xiong} D.~R.,   {Li} S.~K.,  2017, \mn@doi [\apj]
  {10.3847/1538-4357/aa9123}, \href
  {https://ui.adsabs.harvard.edu/abs/2017ApJ...849..161F} {849, 161}

\bibitem[\protect\citeauthoryear{{Fiorucci}, {Ciprini}  \& {Tosti}}{{Fiorucci}
  et~al.}{2004}]{2004A&A...419...25F}
{Fiorucci} M.,  {Ciprini} S.,   {Tosti} G.,  2004, \mn@doi [\aap]
  {10.1051/0004-6361:20034218}, \href
  {https://ui.adsabs.harvard.edu/abs/2004A&A...419...25F} {419, 25}

\bibitem[\protect\citeauthoryear{{Ghisellini}, {Tavecchio}, {Foschini}  \&
  {Ghirlanda}}{{Ghisellini} et~al.}{2011}]{2011MNRAS.414.2674G}
{Ghisellini} G.,  {Tavecchio} F.,  {Foschini} L.,   {Ghirlanda} G.,  2011,
  \mn@doi [\mnras] {10.1111/j.1365-2966.2011.18578.x}, \href
  {https://ui.adsabs.harvard.edu/abs/2011MNRAS.414.2674G} {414, 2674}

\bibitem[\protect\citeauthoryear{{Ghosh}, {Ramsey}, {Sadun}  \&
  {Soundararajaperumal}}{{Ghosh} et~al.}{2000}]{2000ApJS..127...11G}
{Ghosh} K.~K.,  {Ramsey} B.~D.,  {Sadun} A.~C.,   {Soundararajaperumal} S.,
  2000, \mn@doi [\apjs] {10.1086/313313}, \href
  {https://ui.adsabs.harvard.edu/abs/2000ApJS..127...11G} {127, 11}

\bibitem[\protect\citeauthoryear{{Gu} \& {Ai}}{{Gu} \&
  {Ai}}{2011}]{2011A&A...534A..59G}
{Gu} M.~F.,  {Ai} Y.~L.,  2011, \mn@doi [\aap] {10.1051/0004-6361/201117467},
  \href {https://ui.adsabs.harvard.edu/abs/2011A&A...534A..59G} {534, A59}

\bibitem[\protect\citeauthoryear{{Gu}, {Lee}, {Pak}, {Yim}  \& {Fletcher}}{{Gu}
  et~al.}{2006}]{2006A&A...450...39G}
{Gu} M.~F.,  {Lee} C.~U.,  {Pak} S.,  {Yim} H.~S.,   {Fletcher} A.~B.,  2006,
  \mn@doi [\aap] {10.1051/0004-6361:20054271}, \href
  {https://ui.adsabs.harvard.edu/abs/2006A&A...450...39G} {450, 39}

\bibitem[\protect\citeauthoryear{{Gupta}, {Fan}, {Bai}  \& {Wagner}}{{Gupta}
  et~al.}{2008}]{2008AJ....135.1384G}
{Gupta} A.~C.,  {Fan} J.~H.,  {Bai} J.~M.,   {Wagner} S.~J.,  2008, \mn@doi
  [\aj] {10.1088/0004-6256/135/4/1384}, \href
  {https://ui.adsabs.harvard.edu/abs/2008AJ....135.1384G} {135, 1384}

\bibitem[\protect\citeauthoryear{{Heidt} \& {Wagner}}{{Heidt} \&
  {Wagner}}{1996}]{1996A&A...305...42H}
{Heidt} J.,  {Wagner} S.~J.,  1996, \aap, \href
  {https://ui.adsabs.harvard.edu/abs/1996A&A...305...42H} {305, 42}

\bibitem[\protect\citeauthoryear{{Hufnagel} \& {Bregman}}{{Hufnagel} \&
  {Bregman}}{1992}]{1992ApJ...386..473H}
{Hufnagel} B.~R.,  {Bregman} J.~N.,  1992, \mn@doi [\apj] {10.1086/171033},
  \href {https://ui.adsabs.harvard.edu/abs/1992ApJ...386..473H} {386, 473}

\bibitem[\protect\citeauthoryear{{Ikejiri} et~al.,}{{Ikejiri}
  et~al.}{2011}]{2011PASJ...63..639I}
{Ikejiri} Y.,  et~al., 2011, \mn@doi [\pasj] {10.1093/pasj/63.3.327}, \href
  {https://ui.adsabs.harvard.edu/abs/2011PASJ...63..639I} {63, 639}

\bibitem[\protect\citeauthoryear{{Isler}, {Urry}, {Coppi}, {Bailyn}, {Brady},
  {MacPherson}, {Buxton}  \& {Hasan}}{{Isler}
  et~al.}{2017}]{2017ApJ...844..107I}
{Isler} J.~C.,  {Urry} C.~M.,  {Coppi} P.,  {Bailyn} C.,  {Brady} M.,
  {MacPherson} E.,  {Buxton} M.,   {Hasan} I.,  2017, \mn@doi [\apj]
  {10.3847/1538-4357/aa79fc}, \href
  {https://ui.adsabs.harvard.edu/abs/2017ApJ...844..107I} {844, 107}

\bibitem[\protect\citeauthoryear{{Kirk}, {Rieger}  \& {Mastichiadis}}{{Kirk}
  et~al.}{1998}]{1998A&A...333..452K}
{Kirk} J.~G.,  {Rieger} F.~M.,   {Mastichiadis} A.,  1998, \aap, \href
  {https://ui.adsabs.harvard.edu/abs/1998A&A...333..452K} {333, 452}

\bibitem[\protect\citeauthoryear{{Mangalam} \& {Wiita}}{{Mangalam} \&
  {Wiita}}{1993}]{1993ApJ...406..420M}
{Mangalam} A.~V.,  {Wiita} P.~J.,  1993, \mn@doi [\apj] {10.1086/172453}, \href
  {https://ui.adsabs.harvard.edu/abs/1993ApJ...406..420M} {406, 420}

\bibitem[\protect\citeauthoryear{{Mao} \& {Zhang}}{{Mao} \&
  {Zhang}}{2016}]{2016Ap&SS.361..345M}
{Mao} L.,  {Zhang} X.,  2016, \mn@doi [\apss] {10.1007/s10509-016-2934-6},
  \href {https://ui.adsabs.harvard.edu/abs/2016Ap&SS.361..345M} {361, 345}

\bibitem[\protect\citeauthoryear{{Marscher}}{{Marscher}}{2014}]{2014ApJ...780...87M}
{Marscher} A.~P.,  2014, \mn@doi [\apj] {10.1088/0004-637X/780/1/87}, \href
  {https://ui.adsabs.harvard.edu/abs/2014ApJ...780...87M} {780, 87}

\bibitem[\protect\citeauthoryear{{Marscher} \& {Gear}}{{Marscher} \&
  {Gear}}{1985}]{1985ApJ...298..114M}
{Marscher} A.~P.,  {Gear} W.~K.,  1985, \mn@doi [\apj] {10.1086/163592}, \href
  {https://ui.adsabs.harvard.edu/abs/1985ApJ...298..114M} {298, 114}

\bibitem[\protect\citeauthoryear{{Masci} et~al.,}{{Masci}
  et~al.}{2019}]{2019PASP..131a8003M}
{Masci} F.~J.,  et~al., 2019, \mn@doi [\pasp] {10.1088/1538-3873/aae8ac}, \href
  {https://ui.adsabs.harvard.edu/abs/2019PASP..131a8003M} {131, 018003}

\bibitem[\protect\citeauthoryear{{Massaro}, {Maselli}, {Leto}, {Marchegiani},
  {Perri}, {Giommi}  \& {Piranomonte}}{{Massaro}
  et~al.}{2015}]{2015Ap&SS.357...75M}
{Massaro} E.,  {Maselli} A.,  {Leto} C.,  {Marchegiani} P.,  {Perri} M.,
  {Giommi} P.,   {Piranomonte} S.,  2015, \mn@doi [\apss]
  {10.1007/s10509-015-2254-2}, \href
  {https://ui.adsabs.harvard.edu/abs/2015Ap&SS.357...75M} {357, 75}

\bibitem[\protect\citeauthoryear{{Mastichiadis} \& {Kirk}}{{Mastichiadis} \&
  {Kirk}}{2002}]{2002PASA...19..138M}
{Mastichiadis} A.,  {Kirk} J.~G.,  2002, \mn@doi [\pasa] {10.1071/AS01108},
  \href {https://ui.adsabs.harvard.edu/abs/2002PASA...19..138M} {19, 138}

\bibitem[\protect\citeauthoryear{{Padovani} et~al.,}{{Padovani}
  et~al.}{2017}]{2017A&ARv..25....2P}
{Padovani} P.,  et~al., 2017, \mn@doi [\aapr] {10.1007/s00159-017-0102-9},
  \href {https://ui.adsabs.harvard.edu/abs/2017A&ARv..25....2P} {25, 2}

\bibitem[\protect\citeauthoryear{{Rajput}, {Stalin}, {Sahayanathan}, {Rakshit}
  \& {Mandal}}{{Rajput} et~al.}{2019}]{2019MNRAS.486.1781R}
{Rajput} B.,  {Stalin} C.~S.,  {Sahayanathan} S.,  {Rakshit} S.,   {Mandal}
  A.~K.,  2019, \mn@doi [\mnras] {10.1093/mnras/stz941}, \href
  {https://ui.adsabs.harvard.edu/abs/2019MNRAS.486.1781R} {486, 1781}

\bibitem[\protect\citeauthoryear{{Rani} et~al.,}{{Rani}
  et~al.}{2010}]{2010MNRAS.404.1992R}
{Rani} B.,  et~al., 2010, \mn@doi [\mnras] {10.1111/j.1365-2966.2010.16419.x},
  \href {https://ui.adsabs.harvard.edu/abs/2010MNRAS.404.1992R} {404, 1992}

\bibitem[\protect\citeauthoryear{{Safna}, {Stalin}, {Rakshit}  \&
  {Mathew}}{{Safna} et~al.}{2020}]{2020MNRAS.498.3578S}
{Safna} P.~Z.,  {Stalin} C.~S.,  {Rakshit} S.,   {Mathew} B.,  2020, \mn@doi
  [\mnras] {10.1093/mnras/staa2622}, \href
  {https://ui.adsabs.harvard.edu/abs/2020MNRAS.498.3578S} {498, 3578}

\bibitem[\protect\citeauthoryear{{Scarpa}, {Urry}, {Padovani}, {Calzetti}  \&
  {O'Dowd}}{{Scarpa} et~al.}{2000}]{2000ApJ...544..258S}
{Scarpa} R.,  {Urry} C.~M.,  {Padovani} P.,  {Calzetti} D.,   {O'Dowd} M.,
  2000, \mn@doi [\apj] {10.1086/317199}, \href
  {https://ui.adsabs.harvard.edu/abs/2000ApJ...544..258S} {544, 258}

\bibitem[\protect\citeauthoryear{{Schlegel}, {Finkbeiner}  \&
  {Davis}}{{Schlegel} et~al.}{1998}]{1998ApJ...500..525S}
{Schlegel} D.~J.,  {Finkbeiner} D.~P.,   {Davis} M.,  1998, \mn@doi [\apj]
  {10.1086/305772}, \href
  {https://ui.adsabs.harvard.edu/abs/1998ApJ...500..525S} {500, 525}

\bibitem[\protect\citeauthoryear{{Sikora}, {Begelman}  \& {Rees}}{{Sikora}
  et~al.}{1994}]{1994ApJ...421..153S}
{Sikora} M.,  {Begelman} M.~C.,   {Rees} M.~J.,  1994, \mn@doi [\apj]
  {10.1086/173633}, \href
  {https://ui.adsabs.harvard.edu/abs/1994ApJ...421..153S} {421, 153}

\bibitem[\protect\citeauthoryear{{Silverman}}{{Silverman}}{1986}]{1986desd.book.....S}
{Silverman} B.~W.,  1986, {Density estimation for statistics and data analysis}

\bibitem[\protect\citeauthoryear{{Urry} \& {Mushotzky}}{{Urry} \&
  {Mushotzky}}{1982}]{1982ApJ...253...38U}
{Urry} C.~M.,  {Mushotzky} R.~F.,  1982, \mn@doi [\apj] {10.1086/159607}, \href
  {https://ui.adsabs.harvard.edu/abs/1982ApJ...253...38U} {253, 38}

\bibitem[\protect\citeauthoryear{{Urry} \& {Padovani}}{{Urry} \&
  {Padovani}}{1995}]{1995PASP..107..803U}
{Urry} C.~M.,  {Padovani} P.,  1995, \mn@doi [\pasp] {10.1086/133630}, \href
  {https://ui.adsabs.harvard.edu/abs/1995PASP..107..803U} {107, 803}

\bibitem[\protect\citeauthoryear{{Vagnetti}, {Trevese}  \& {Nesci}}{{Vagnetti}
  et~al.}{2003}]{2003ApJ...590..123V}
{Vagnetti} F.,  {Trevese} D.,   {Nesci} R.,  2003, \mn@doi [\apj]
  {10.1086/374889}, \href
  {https://ui.adsabs.harvard.edu/abs/2003ApJ...590..123V} {590, 123}

\bibitem[\protect\citeauthoryear{{Van Roestel} et~al.,}{{Van Roestel}
  et~al.}{2021}]{2021AJ....161..267V}
{Van Roestel} J.,  et~al., 2021, \mn@doi [\aj] {10.3847/1538-3881/abe853},
  \href {https://ui.adsabs.harvard.edu/abs/2021AJ....161..267V} {161, 267}

\bibitem[\protect\citeauthoryear{{Vaughan}, {Edelson}, {Warwick}  \&
  {Uttley}}{{Vaughan} et~al.}{2003}]{2003MNRAS.345.1271V}
{Vaughan} S.,  {Edelson} R.,  {Warwick} R.~S.,   {Uttley} P.,  2003, \mn@doi
  [\mnras] {10.1046/j.1365-2966.2003.07042.x}, \href
  {https://ui.adsabs.harvard.edu/abs/2003MNRAS.345.1271V} {345, 1271}

\bibitem[\protect\citeauthoryear{{Villata} et~al.,}{{Villata}
  et~al.}{2004}]{2004A&A...421..103V}
{Villata} M.,  et~al., 2004, \mn@doi [\aap] {10.1051/0004-6361:20035895}, \href
  {https://ui.adsabs.harvard.edu/abs/2004A&A...421..103V} {421, 103}

\bibitem[\protect\citeauthoryear{{Wagner} \& {Witzel}}{{Wagner} \&
  {Witzel}}{1995}]{1995ARA&A..33..163W}
{Wagner} S.~J.,  {Witzel} A.,  1995, \mn@doi [\araa]
  {10.1146/annurev.aa.33.090195.001115}, \href
  {https://ui.adsabs.harvard.edu/abs/1995ARA&A..33..163W} {33, 163}

\bibitem[\protect\citeauthoryear{{Wiita}, {Miller}, {Carini}  \&
  {Rosen}}{{Wiita} et~al.}{1991}]{1991sepa.conf..557W}
{Wiita} P.~J.,  {Miller} H.~R.,  {Carini} M.~T.,   {Rosen} A.,  1991, in
  {Bertout} C.,  {Collin-Souffrin} S.,   {Lasota} J.~P.,  eds, IAU Colloq. 129:
  The 6th Institute d'Astrophysique de Paris (IAP) Meeting: Structure and
  Emission Properties of Accretion Disks. p.~557

\bibitem[\protect\citeauthoryear{{Zhang}, {Zhou}, {Zhao}  \& {Dai}}{{Zhang}
  et~al.}{2015}]{2015RAA....15.1784Z}
{Zhang} B.-K.,  {Zhou} X.-S.,  {Zhao} X.-Y.,   {Dai} B.-Z.,  2015, \mn@doi
  [Research in Astronomy and Astrophysics] {10.1088/1674-4527/15/11/002}, \href
  {https://ui.adsabs.harvard.edu/abs/2015RAA....15.1784Z} {15, 1784}

\makeatother
\end{thebibliography}




\appendix




\bsp	
\label{lastpage}
\end{document}